\documentclass[prc,amsmath,twocolumn,superscriptaddress]{revtex4} 

\usepackage{dcolumn}
\usepackage{graphicx}
\usepackage{color}
\usepackage{natbib}

\def\beq{\begin{equation}}
\def\eeq{\end{equation}}
\def\bea{\begin{eqnarray}}
\def\eea{\end{eqnarray}}

\renewcommand*{\eqref}[1]{Eq.~(\ref{eq:#1})}
\newcommand*{\eqlab}[1]{\label{eq:#1}}
\newcommand*{\figref}[1]{Fig.~\ref{fig:#1}}
\newcommand*{\figlab}[1]{\label{fig:#1}}

\newcommand{\Omit}[1]{}

\begin{document}

\title{Improved flux limits for neutrinos with energies above 10$^{22}$\,eV
from observations with the Westerbork Synthesis Radio Telescope}

 \author{O. Scholten} 
 \affiliation{Kernfysisch Versneller Instituut, University of Groningen, 9747 AA, Groningen, The Netherlands}
\author{S. Buitink} 
\affiliation{ Department of Astrophysics, IMAPP, Radboud University, 6500 GL
Nijmegen, The Netherlands}
 \affiliation{Lawrence Berkeley National Laboratory, Berkeley, California 94720, USA}
 \author{J. Bacelar} 
 \affiliation{ASML Netherlands BV, P.O.Box 324, 5500 AH Veldhoven, The  Netherlands}
 \author{R. Braun} 
 \affiliation{CSIRO-ATNF, P.O.Box 76, Epping NSW 1710, Australia}
 \author{A.G. de Bruyn} 
 \affiliation{Kapteyn Institute, University of Groningen, 9747 AA, Groningen, The Netherlands}
 \affiliation{ASTRON, 7990 AA Dwingeloo, The Netherlands}
 \author{H. Falcke} 
 \affiliation{ Department of Astrophysics, IMAPP, Radboud University, 6500 GL Nijmegen, The Netherlands}
 \author{K. Singh} 
 \affiliation{Kernfysisch Versneller Instituut, University of Groningen, 9747 AA, Groningen, The Netherlands}
 \author{B. Stappers} 
 \affiliation{Jodrell Bank Centre for Astrophysics,
School of Physics and Astronomy, The University of Manchester, Manchester M13
9PL, UK }
 \author{R.G. Strom} 
  \affiliation{ASTRON, 7990 AA Dwingeloo, The Netherlands}
 \affiliation{Astronomical Institute `A. Pannekoek', University of Amsterdam, 1098 SJ, The Netherlands }
 \author{R. al Yahyaoui} 
 \affiliation{Kernfysisch Versneller Instituut, University of Groningen, 9747 AA, Groningen, The Netherlands}


\begin{abstract}
Particle cascades initiated by ultra-high energy (UHE) neutrinos in the lunar
regolith will emit an electromagnetic pulse with a time duration of the order of
nano seconds through a process known as the Askaryan effect. It has been shown
that in an observing window around 150~MHz there is a maximum chance for
detecting this radiation with radio telescopes commonly used in astronomy. In 50
hours of observation time with the Westerbork Synthesis Radio Telescope array we
have set a new limit on the flux of neutrinos, summed over all flavors, with
energies in excess of $4\times10^{22}$~eV.
\end{abstract}
\maketitle

\section{Introduction}

At high energies the spectrum of cosmic rays follows a power law distribution
extending up to extremely large energies. At the Pierre Auger Observatory cosmic
rays have been observed~\cite{A08} with energies in excess of $\sim 10^{20}$~eV.
Above the Greisen-Zatsepin-Kuzmin (GZK) energy of $6\times 10^{19}$~eV, cosmic
rays can interact with the photons of the cosmic microwave background to produce
pions~\cite{G66,ZK66} which carry a sizable fraction of the original energy of
the cosmic ray. Charged pions decay and produce neutrinos and one thus may expect
the presence of neutrinos with energies in excess of the GZK energy. Recently at
the Pierre Auger Observatory a steepening of the slope in the cosmic ray spectrum
has been observed at the GZK energy~\cite{A08} which can be regarded as a
consequence of the GZK effect.

Since neutrinos are chargeless they will propagate in a straight line with
negligible energy loss from the location where they have been created to the
observer, thus carrying direct information on their source. These sources could
be the aforementioned processes related to the GZK effect or, more exotically,
decaying supermassive dark-matter particles or topological defects. This last
class of models is referred to as top-down (see \citet{s04} for a review).

Because of their small interaction cross section, the detection of cosmic
neutrinos calls for extremely large detectors. At GeV energies their cross
section is so minute that at a flux given by the Waxman-Bahcall
estimate~\cite{Waxman:1998yy} of a few tens of neutrinos per km$^2$ per year one
needs to employ km$^3$-scale detectors~\cite{icecube,KM3NET}. At higher neutrino
energies the reaction cross section increases. However, their flux is expected to
fall even faster and one needs even larger detection volumes. These can be
obtained by observing large detector masses from a distance. The ANITA balloon
mission~\cite{anita08} monitors an area of a million km$^2$ of South Pole ice and
the FORTE satellite~\cite{forte} can pick up radio signals coming from the
Greenland ice mass. Alternatively the Pierre Auger Observatory can distinguish
cosmic ray induced air showers from neutrino induced cascades at very high zenith
angles~\cite{Abraham:2007rj}.

The Moon offers an even larger natural detector volume. In the interaction of an
UHE neutrino about 20\% of the energy of the neutrino is converted into a cascade
of high-energy particles, called the hadronic shower. Due to the electromagnetic
component of this shower the electrons in the material are swept out from the
atom to become part of the shower. The shower thus has an excess of negative
charge moving with relativistic velocities through a material with an index of
refraction considerably different from unity resulting in the emission of
Cherenkov radiation. Since the lateral side of the shower has a dimension of the
order of 10~cm, the radio emission is coherent for wavelength of this magnitude
and larger or for frequencies up to $\sim 3$ GHz. The emission of coherent
Cherenkov radiation in such a process is known as the Askaryan effect~\cite{a62}.
This emission mechanism has been experimentally verified at
accelerators~\cite{s01} and extensive calculations have been performed to
quantify the effect~\cite{zhs92,az97}.

For showers in the lunar regolith, the top layer of the Moon consisting of dust
and small rocks, its properties are important. Much is known from samples brought
from the Moon~\cite{os75}. The average index of refraction is $n=1.8$ and the
attenuation length is $\lambda_{r}=(9/\nu[\mathrm{GHz}])$~m for radio waves. The
thickness of the regolith is known to vary over the lunar surface. At some depth
there is a (probably smooth) transition to solid rock, for which the density is
about twice that of the regolith.

At the highest energies the emitted pulse from the Moon can be observed at Earth
with radio telescopes~\cite{dz89}. The first experiments in this direction were
carried out with the Parkes telescope~\cite{parkes}, later followed by
others~\cite{glue,KALYAZIN}. A recent project is LUNASKA~\cite{James:2009rc} that
is currently performing lunar Cherenkov measurements with the Australia Telescope
Compact Array with a 600 MHz bandwidth at 1.2-1.8~GHz. These observations are all
performed at relatively high frequencies (2~GHz) where the emission is strongest.
At lower frequencies~\cite{Fal03} the angular spread of the emission around the
Cherenkov angle increases due to finite source effects~\cite{scholten}. When the
wavelength is similar to the longitudinal extent of the shower in the lunar rock,
a few meters, the angular spread is close to isotropic and the probability of
detecting the radio pulse is largest~\cite{scholten}. In our observations we
exploit this optimal frequency range around 150 MHz using the Westerbork
Synthesis Radio Telescope (WSRT).

\section{Observations with the WSRT}
\label{sec:detection}

The WSRT consists of an array of 14 parabolic antennas of 25~m diameter on a
2.7~km east-west line. In the observations we use the Low Frequency Front Ends
(LFFEs) which cover the frequency range 115--180~MHz with full polarization
sensitivity. The Pulsar Machine II (PuMa II) backend~\cite{kss08} can record a
maximum bandwidth of 160~MHz, sampled as 8 subbands of 20~MHz each. Only 11 of
the 12 equally spaced WSRT dishes are used for this experiment which means that
when the telescopes are added in phase the resultant beam on the sky is a fan
beam\cite{Janssen:2009ya}. The phases required to add the dishes coherently are
determined by observations of a known calibrator source, which at these
frequencies is Cassiopeia A. Adjusting the phase relations between the 8 subbands
they can be pointed to any location within the primary beam of the 25\,m dish. We
therefore choose to use 4 frequency bands centered at frequencies of 123, 137,
151 and 165 MHz repeated for two different look directions or beams, aimed at
different sides of the Moon, each covering about one third of the lunar surface.
This increases the effective aperture and creates the possibility of an
anti-coincidence trigger since a lunar Cherenkov pulse should only be visible in
one of the two beams. Because of overlap in the band width of the sub-bands the
total bandwidth per beam is 65~MHz. The system has a real time automatic gain
control (AGC) system, that stabilizes the average gain of the output signal. The
time series data is recorded for each subband with a sampling frequency of 40
MHz. The timeseries nature of the data and the lack of calibrated amplitude
calibrator signal means that an accurate absolute calibration of the signal
strength is not possible and we therefore express the pulse strength in terms of
the power of the background noise $\sigma^2$, the System Equivalent Flux Density.
Averaged over the frequency range under consideration this amounts to
$\sigma^2=400$~Jy where care has been taken not to have the galactic plane in the
field of view.

The data is processed in blocks of 0.1~s, where each block is divided in 200
traces of 20,000 time samples. The data analysis is performed in three steps.

First, the narrow band Radio Frequency Interference (RFI) is filtered from the
data. The Fourier transforms of the 200 time traces in one block are added to
suppress statistical fluctuations and fitted by a 9$^{th}$ order polynomial. All
frequency channels exceeding the fit by 50\% or more are marked as RFI lines and
set equal to zero. This is done for all 8 frequency bands and both polarizations.
The loss of band width due to this filtering is less than 2\%.

Second, the dispersion due to the ionosphere of the Earth is corrected for. The
vertical total electron content (TEC) values of the ionosphere are provided by
the DLR Institut f\"ur Kommunikation und Navigation.
These values are corrected for the elevation of the Moon to obtain the
slanted-TEC (STEC) value. Because of variations of the ionosphere on short
timescales and uncertainties in the vertical density profile we allow for an
inaccuracy in the determined STEC value. From the STEC value a frequency
dependent phase is calculated which is used in the inverse fourier transform to
obtain dispersion and RFI corrected time traces. The error in the STEC value may
result in an increased time width of the pulses and an offset between the arrival
times of pulses in different frequency bands.

Third, the spectrum is searched for strong pulses with large bandwidth. From the
corrected time spectra a 5-time-sample-summed relative-power spectrum is
constructed,
\begin{equation}
P_5=\frac{\displaystyle\sum_{\mathrm{5\ samples}} P_x}{\bigg<\displaystyle\sum_{\mathrm{5\ samples}} P_x\bigg>}+
\frac{\displaystyle\sum_{\mathrm{5\ samples}} P_y}{\bigg<\displaystyle\sum_{\mathrm{5\ samples}} P_y\bigg>},
\label{P5}
\end{equation}
where the averaging is done over one time trace (20,000 time samples), and $x$
and $y$ denote the two polarizations. By summing over 5 time samples one allows
for an error in the STEC value and for the spreading of a bandwidth-limited
Nyquist-sampled pulse. The $P_5$-spectra are searched for values exceeding 5. The
first and last 250 time samples, corresponding to 0.25\% of the observation time,
are excluded from this search since the RFI noise is not properly eliminated from
these parts of the spectrum. A trigger is generated when in all four frequency
bands in the same beam a value $P_5>5$ is found, within a maximum time offset of
$ \Delta t =\mathrm{STEC} \times 40 \frac{(100MHz)^2}{\nu_1\nu_2}
\frac{\nu_{2}^2-\nu_{1}^2}{\nu_1\nu_2}$~ns, to account for a 30\% error on the
STEC value (in units of $TECU=10^{16} el/m^2$). No search is done for a second
pulse in the same trace. The value $S$ is defined as the sum over the maximum
$P_5$ values in the 4 frequency bands,
\begin{equation}
S=\sum_{\mathrm{4\ bands}} P_5 \;. \eqlab{S}
\end{equation}

\begin{figure}
\centering
\includegraphics[width=0.9\linewidth,viewport=4 4 515 355,clip]{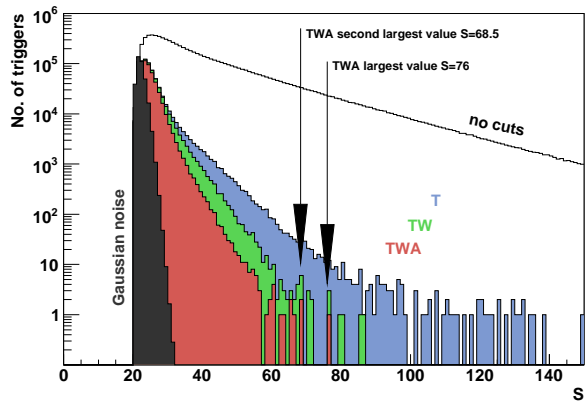}
\caption{[color online]Distribution of $S$ for increasingly stronger cuts.
The top curve represents the distribution of raw
triggers where subsequently the, T, TW, and the TWA cuts are applied.}
\label{fig:distribution}
\end{figure}

\figref{distribution} shows the distribution of $S$ for the triggered events. In
a subsequent analysis additional constraints are imposed.

{\bf Timer signal (T)} The data contains short strong pulses that repeat at a
regular interval with a frequency of 102.4~s$^{-1}$ suggesting an instrumental
origin. Cutting out the time intervals in which these pulses occur corresponds to
a loss of $\sim$10\% of observation time.

{\bf $P_5$ width (W)} The number of consecutive $P_5$ values exceeding the
threshold should be limited for a real lunar pulse. On the basis of simulations
the width cut is set at $W<12$.

{\bf Anti-coincidence (A)} A lunar pulse should be visible in only one of the two
beams. An anti-coincidence trigger is set up by excluding events in which a pulse
was found in both beams in the same time trace.

\figref{distribution} shows distributions of $S$ after application of only the
timer cut, the timer and width cut (TW), and a combination of all cuts (TWA). The
line enclosing the black area corresponds to the expected number of triggers for
pure Gaussian noise. After all cuts have been applied the number of triggers for
which $S>23$ is a factor of 3-4 higher than the amount of triggers expected for
Gaussian noise. The largest remaining pulse has a value of $S=76$ which is used
for setting a limit on the flux of UHE neutrinos. This is a factor 3 larger than
expected for statistical noise which reflects in less stringent limits that
estimated in Ref.~\cite{scholten}. A full account of the analysis will be
presented in Ref.~\cite{FullPaper}.

A closer examination of the pulse with $S=76$ excludes the possibility of it
being a lunar pulse. More than half of the total power is received in one of the
four frequency bands, while the other bands triggered on pulses that are of
comparable size as background pulses. Moreover, the same large peak is present in
the corresponding frequency band of the other beam. The anti-coincidence cut did
not remove this event since in that beam not all other bands had a trigger,
indicating that our limit is rather conservative.

\section{Detection Efficiency}

To determine our detection efficiency, pulses with a time-duration less than the
sampling time are dispersed corresponding to a typical TEC value, simTEC=12. The
amplitudes are Nyquist-sampled with a random time offset, rounded off to the
nearest integer within the dynamic range, and added to the raw data (considered
as background). The detection efficiency (DE) is the fraction of inserted pulses
that is retrieved after applying the trigger conditions and the cuts that are
used in the analysis. For the simulations we have inserted 1000 pulses in various
10 second segments of data.

\begin{figure}
\centering
\includegraphics[width=0.7\linewidth,viewport=83 120 570 595,clip]{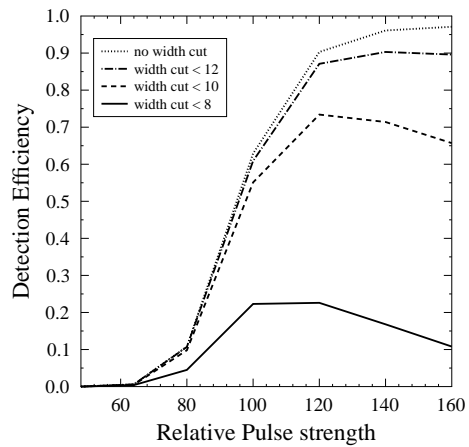}
\caption{ The detection efficiency is shown as a function of pulse
strength for various settings of the trigger conditions as discussed in the text.
The dotted curve indicates the efficiency when no width cut is applied.
}
\label{fig:farad}
\end{figure}

\figref{farad} shows the DE for inserted pulses of strength varying from $S_i=50$
to $S_i=160$. The pulse from the Moon is polarized in the direction of the
shower. In the simulations we have therefore taken into account the Faraday
rotation of the polarization vector. The de-dispersion is done with STEC=10,
different from simTEC, to simulate an error in the STEC value. The lines in
\figref{farad} show the DE for recovering pulses with strength exceeding
$S_{th}=77$. Due to interference from the background the recovered pulse strength
differs from the input value $S_i$. The dotted lines show the DE without any
width cut applied. Dash-dotted (dashed, solid) lines represents the DE with width
cut $W< 12$, ($W< 10$, $W < 8$) respectively. Clearly the $W< 12$ is close to the
optimum and is selected for further analysis.

After correcting for hardware and software failures, excluding four observation
runs which have exceptionally large numbers of raw triggers, correcting for dead
time due to double hits, and the effect of timer signal correction, we have 46.7
hours of dual-beam observation time left. Each beam covers about a third of the
lunar area.

\section{Results}

In 46.7 hours of observation time no pulses from the Moon were found for which
$S>77$. To convert this into a probability for not observing a pulse from the
Moon we use \figref{farad} which shows that only pulses with strength in excess
of $S_i>120$ have a large DE of 87.5\%. For smaller pulses from the Moon, due to
interference with background, the DE drops rapidly which will give rise to less
constraining limits.

\begin{figure}
\centering
\includegraphics[width=0.8\linewidth,viewport=50 150 550 700,clip]{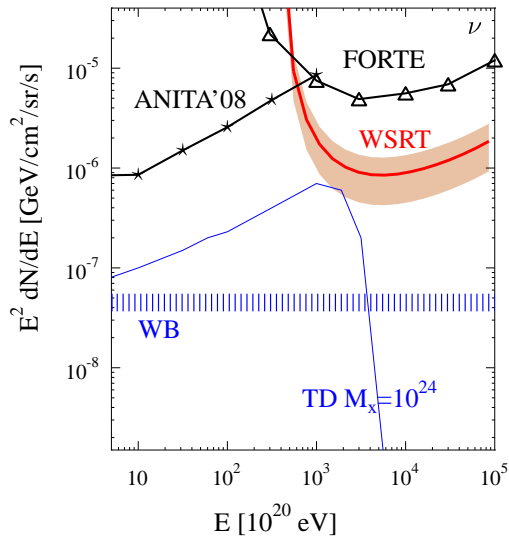}
\caption{[color online] The currently established neutrino flux limit is compared to
the limits set by ANITA~\cite{anita08} and FORTE~\cite{forte}
as well as the Waxman-Bahcall flux~\cite{Waxman:1998yy} and a TD model prediction~\cite{ps96}.
The band shows the systematic error.}
\figlab{wsrtlimits}
\end{figure}

With the applied constraints we thus have a 87.5\% probability to observe an
emitted pulse with $S_i>120$ corresponding to 120$\sigma^2\times 5=240$~kJy. The
lack of pulses of a certain power implies a limit on the flux of neutrinos. This
is not a simple algebraic relation since attenuation of the signal in the Moon,
transmission at the lunar surface and angle with respect to the direction of the
neutrino all affect the observed pulse strength. On the basis of the simulations
which are described in detail in Ref.~\cite{scholten} the 90\% confidence limit
flux limit shown in \figref{wsrtlimits} has determined. In arriving at this the
model-independent procedure described in Ref.~\cite{forte} has been followed.

In arriving at this limit the same assumptions have been made as in
Ref.~\cite{scholten}, in particular that the neutrino cross sections equal the
predictions given in Ref.~\cite{Gan00}. The estimated systematic errors on the
acceptance simulation~\cite{scholten} are due to uncertainties in the: density
(10\% in threshold energy), attenuation length (40\% in flux), and stopping power
of the regolith, (20\% in flux). The error in the moon coverage of the two beams
is estimated at 20\%. Adding these errors in quadrature gives a systematic error
on the flux of 50\% as indicated in \figref{wsrtlimits}.

The present limits in the UHE region have been set by ANITA~\cite{anita08} and
FORTE~\cite{forte}. On the basis of the present observations the flux limits have
been improved by an order of magnitude at the highest energies. The new limit is
still well above the Waxman-Bahcall limit~\cite{Waxman:1998yy} but borders on the
predictions of a top-down model~\cite{ps96} for exotic particles of mass
$M_X=10^{24}$~eV.

Presently similar techniques as used in the WSRT observations are implemented for
observations with the LOFAR radio telescope (presently being rolled out) and the
future SKA telescope. With the latter we should be able to observe neutrinos
associated with the GZK effect~\cite{Sch09}.

\begin{acknowledgments}
This work was performed as part of the research programs of the Stichting voor
Fundamenteel Onderzoek der Materie (FOM) and of ASTRON, both with financial
support from the Nederlandse Organisatie voor Wetenschappelijk Onderzoek (NWO).
\end{acknowledgments}


\end{document}